# A unified descriptor framework for hydrogen storage capacity and equilibrium pressure in interstitial hydrides

*Seong-Hoon Jang*[*1, 2], *Di Zhang*[1,3], *Xue Jia*[1], *Hung Ba Tran*[1], *Linda Zhang*[1,3], *Ryuhei Sato*[4], *Yusuke Hashimoto*[3], *Yusuke Ohashi*[5], *Toyoto Sato*[5], *Kiyoe Konno*[6], *Shin-ichi Orimo*[*1, 5], *and Hao Li*[*1]

[1] Advanced Institute for Materials Research (WPI-AIMR), Tohoku University, Sendai 980-8577, Japan

[2] Unprecedented-scale Data Analytics Center, Tohoku University, Sendai 980-8578, Japan

[3] Frontier Research Institute for Interdisciplinary Sciences (FRIS), Tohoku University, Sendai 980-8577, Japan

[4] Department of Materials Engineering, The University of Tokyo, Tokyo 113-8656, Japan

[5] Institute for Materials Research, Tohoku University, Sendai, 980-8577

[6] Institute of Fluid Science, Tohoku University, Sendai, 980-8577, Japan

* Corresponding authors:

jang.seonghoon.b4@tohoku.ac.jp (S.-H. Jang)

shin-ichi.orimo.a6@tohoku.ac.jp (S. Orimo)

li.hao.b8@tohoku.ac.jp (H. Li)

ABSTRACT

Hydrogen is a promising energy carrier, yet its practical deployment is limited by the lack of storage materials that simultaneously achieve high storage capacity ($w$) and practical equilibrium pressure at room temperature ($P_{\text{eq,RT}}$). Interstitial metal hydrides offer fast kinetics and favorable thermodynamics (high $P_{\text{eq,RT}}$) but suffer from intrinsically low $w$. Here, we establish a physically interpretable, data-driven framework to uncover descriptor-property relationships in interstitial hydrides using a curated database of pressure-composition-temperature measurements (Digital Hydrogen Platform, *DigHyd*) and white-box symbolic regression. **Strikingly, the analysis reveals a clear separation of governing mechanisms, in which $w$ is governed by geometric and lattice conditions, captured by the average atomic radius ($\langle r_M \rangle$) and average thermal conductivity ($\langle \kappa \rangle$), with an optimal regime of $\langle r_M \rangle \sim 1.47$ Å and relatively low $\langle \kappa \rangle$. In contrast, $P_{\text{eq,RT}}$ is governed by elastic properties, captured by the average shear modulus ($\langle G \rangle$) and average Poisson's ratio ($\langle \nu \rangle$), reflecting the role of lattice rigidity and mechanical compliance.** These relationships are translated into compositional optimization pathways that follow the descriptor trends above, enabling the design of candidate materials with enhanced $w$ under practical equilibrium conditions ($P_{\text{eq,RT}} \sim 0.1$ MPa). This work establishes a general, interpretable strategy for physics-informed design of energy materials systems.

**Introduction**

Hydrogen is widely regarded as a key enabler for carbon-neutral energy systems, primarily due to its high energy density by mass and the absence of carbon emissions upon use.[1, 2] Despite these advantages, its widespread application in fuel cells and related technologies remains constrained by the lack of storage solutions that are simultaneously compact, safe, and reversible.[3] Among the various approaches proposed to address this challenge, solid-state hydrogen storage in metal hydrides has attracted considerable attention because of its high volumetric density, cyclability, and compatibility with engineered systems.[4-6] Metal hydrides, including representative systems such as $MgH_2$, $Mg_2NiH_4$, $FeTiH_2$, $PdH_{0.6}$, and $LaNi_5H_6$, span a wide thermodynamic range depending on their bonding characteristics.[7-15] Saline-type hydrides composed of light elements (*e.g.,* $MgH_2$) can achieve high gravimetric capacities but typically require elevated temperatures for hydrogen release,[13, 16] whereas interstitial hydrides based on transition or heavier metals (*e.g.,* $LaNi_5H_6$) exhibit fast kinetics and favorable equilibrium pressures, albeit with intrinsically limited hydrogen storage capacity.[11]

Despite decades of extensive investigation, the compositional landscape of hydride-forming alloys remains far from fully explored. While a vast number of binary and multicomponent systems are theoretically accessible, only a limited fraction has been experimentally synthesized and evaluated. **This challenge is further exacerbated by the absence of predictive frameworks that are both quantitatively accurate and physically interpretable**, which hinders rational materials design. Although recent machine learning approaches have shown promise in accelerating property prediction, they often rely on relatively small or inconsistently curated datasets and employ black-box models that provide limited insight into the underlying physicochemical mechanisms. To overcome these limitations, we previously developed the Digital Hydrogen Platform (*DigHyd*: www.dighyd.org), a curated database constructed through large-scale extraction of experimental pressure-composition-temperature (PCT) data from the literature.[17, 18] Building on this dataset, symbolic regression was performed using a white-box modeling approach, enabling the construction of explicit relationships between materials descriptors and key hydrogen storage metrics, namely gravimetric capacity ($w$) and equilibrium pressure at room temperature ($P_{eq,RT}$).[19, 20] This approach identified a compact set of physically meaningful descriptors, including contributions from atomic mass, electronic structure, and packing characteristics, which govern hydrogen storage behavior. In particular, systems containing light elements were found to favor higher hydrogen capacity, whereas electronic and structural descriptors play a dominant role in determining equilibrium pressure. Within this descriptor space, beryllium (Be)-containing alloys emerged as promising candidates; however, their practical applicability is severely limited by toxicity and associated handling constraints.[21]

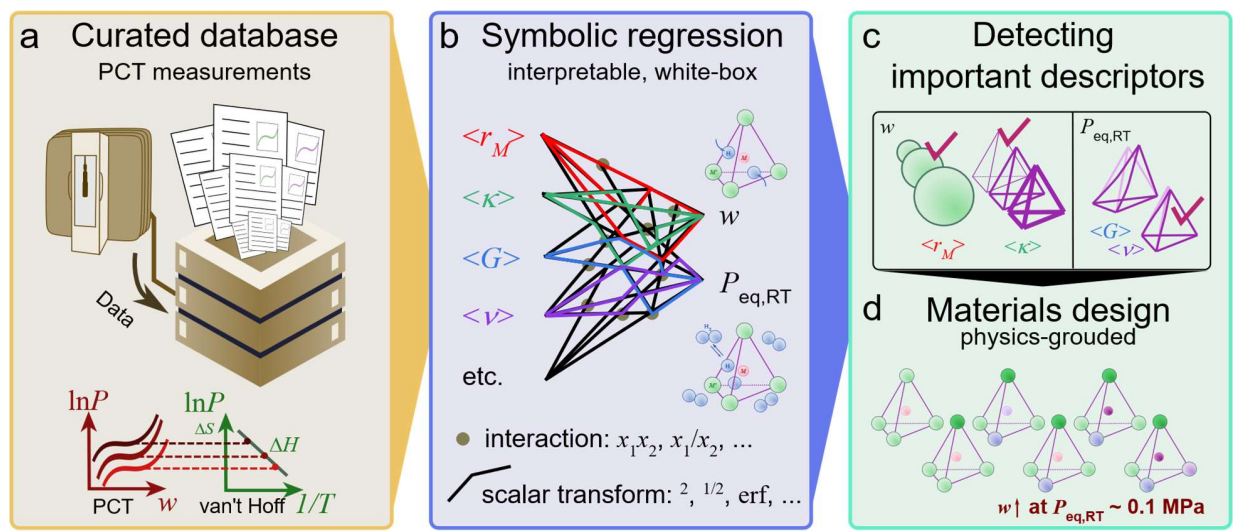

**Figure 1** Workflow of the present study for physics-informed materials design of interstitial hydrides. (a) Construction of a curated database of pressure-composition-temperature (PCT) measurements (*DigHyd*). (b) Development of interpretable, white-box symbolic regression models linking materials descriptors to hydrogen storage properties, including gravimetric capacity ($w$) and equilibrium pressure at room temperature ($P_{eq,RT}$). (c) Identification of key descriptors governing $w$ and $P_{eq,RT}$ within the interstitial hydride space. (d) Translation of descriptor-property relationships into materials design guidelines, enabling the optimization of $w$ under practical equilibrium pressure conditions ($P_{eq,RT} \sim 0.1$ MPa).

These considerations motivate a shift toward more practical materials systems, as schematically illustrated in **Figure 1**. **Figure 1a** shows the construction of a curated database of PCT measurements, *DigHyd*, which serves as the foundation for this study. Building on this dataset, **Figure 1b** presents the development of interpretable, white-box symbolic regression models that link materials descriptors to key hydrogen storage properties, $w$ and $P_{eq,RT}$. Focusing on interstitial hydrides, we aim to identify the key descriptors governing both $w$ and $P_{eq,RT}$ by extracting physically meaningful descriptors from the symbolic models (see **Figure 1c**). Finally, as shown in **Figure 1d**, **these descriptor-property relationships are translated into materials design guidelines that enable enhanced $w$ under practical operating conditions.** In particular, we target equilibrium pressures near ambient conditions, $P_{\text{eq,RT}} \sim 0.1 \text{MPa}$, corresponding in this study to the window of $-1.5 \ \log_{10}[\text{MPa}] < \log_{10} P_{\text{eq,RT}} < -0.5 \ \log_{10}[\text{MPa}]$.

**Results**

**Data Curation and Feature Construction for Interstitial Hydrides**

Often, PCT experiments involve multi-phase materials, which can introduce unintended noise in regression modeling. To mitigate this issue, we filtered the dataset and retained only single-phase or near-single-phase cases. Here, "near-single-phase" refers to materials in which a single phase accounts for more than 80 wt. % of the total. As a result, the total number of entry is 706. However, not all entries include multi-temperature measurements required to determine $P_{\text{eq,RT}}$. Consequently, the number of data point ($n_{\text{data}}$) used for modeling $w$ and $P_{\text{eq,RT}}$ is 706 and 299, respectively.

**Figure 2a** shows the distribution of data points in the two-dimensional $w - P_{\text{eq,RT}}$ materials map for cases where multi-temperature PCT data are available. Most structural classes with different parent structures, such as Laves (C14), LaNi$_5$, LaMgNi$_4$, and TiFe, are located in the low-$w$ region ($w < 2.5$ %), whereas the BCC class extends into the higher-capacity region ($2 < w < 5$ %). However, BCC metals and alloys generally exhibit two distinct plateaus in their PCT curves, with the lower plateau typically located far below atmospheric pressure at room temperature, thereby limiting the practical accessibility of their full $w$.[22] **Figure 2b** presents $n_{\text{data}}$ for each reported class. Notably, three classes, Laves (C14), LaNi$_5$, and BCC, out of 14 classes account for $n_{\text{data}} = 198$, corresponding to 66.2 % of the dataset with multi-temperature PCT measurements ($n_{\text{data}} = 299$), which is close to the Pareto principle.

For symbolic regression modeling, features were extracted as candidate descriptors for $w$ and $P_{\text{eq,RT}}$. The full list of features and their denotations is provided in **Table 1**; hereafter, the denotations are omitted for simplicity. A total of 57 features were constructed, comprising 18 chemical, physical, and structural features applied to their averaged values $\langle \cdots \rangle$, standard deviations $\sigma(\cdots)$, and skewness $r(\cdots)$, along with three additional compositional features ($R_{\text{Tr}}$, $R_{\text{Tr(IV)}}$, and $R_{\text{Tr+R}}$). For example, for a compound $A_a B_b C_c$, the averaged atomic mass is defined as $\langle M \rangle = (aM_a + bM_b + cM_c)/(a + b + c)$, where $M_x$ denotes the atomic mass of element $x$. Structural descriptors such as $\Omega$, $\Omega_\sigma$, and $V_p$ require the definition of coordination polyhedra $XY_n$ ($X, Y = A, B,$ and $C$ crystallographic sites; $X$-centered). The maximum cutoff distance for metal-metal pairs was set to 3.5 Å, corresponding to the shoulder of the first peak in the radial distribution function averaged across crystal structures (see the section **"Averaged Radial Distribution Function of Metal-alloy Parent Structures and the Construction of $XY_n$ Polyhedra"** in the **Supplementary Information**). In addition, the heatmap of Pearson correlation coefficients ($r_{col}$) among all feature pairs is provided in the section **"Pearson Correlation Heatmap"** in the **Supplementary Information**.

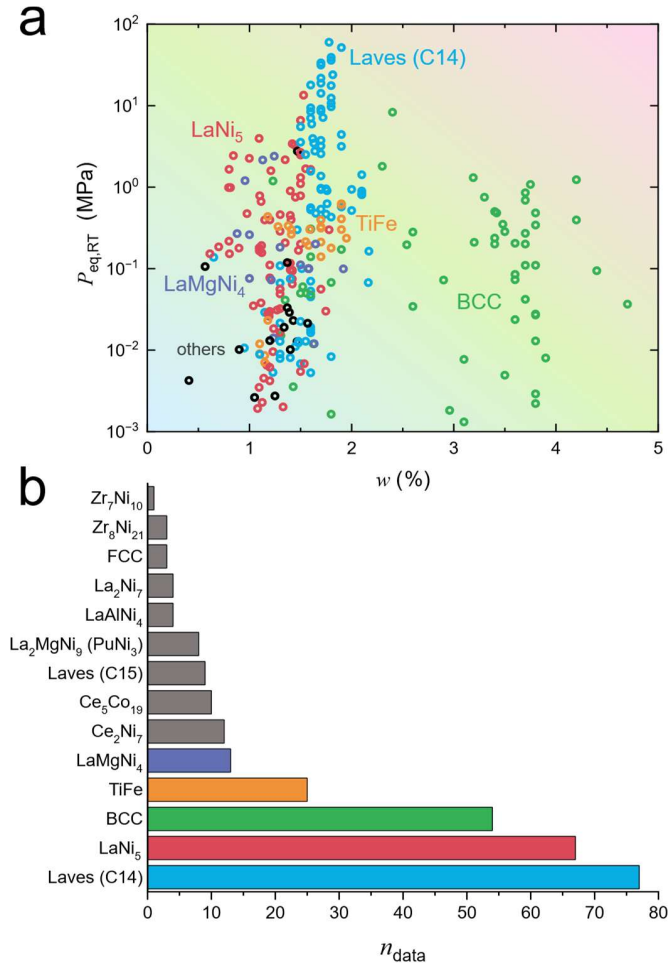

**Figure 2** Data distribution and feature construction for interstitial hydrides. (a) Distribution of data points in the $w - P_{eq,RT}$ map for entries with multi-temperature PCT data. (b) Number of data points ($n_{data}$) for each structural class.

**Table 1** Chemical, physical, structural, and compositional properties of constituent elements of interstitial metal hydrides for regression models, given as features for symbolic regression modeling.

| Properties | Description | Unit |
|---|---|---|
| $\chi - \chi_{\mathrm{H}}$ | Difference in electronegativity between metal atom and hydrogen | - |
| $M$ | Atomic mass | $\mathrm{g \cdot mol^{-1}}$ |
| $n_{ve}$ | The number of valence electrons. For $s/p$-block and $d/f$-block, electrons in the shells of $n_p s$ and $n_p p$, and in those of $n_p s$ and $(n_p - 1)d$ are counted, respectively; $n_p$ is the highest principal quantum number. | - |
| $r_M$ | Metallic radius | Å |
| $\rho$ | Density | $\mathrm{g \cdot cm^{-3}}$ |
| $\rho_{\mathrm{mol}}$ | Molar density, defined as $\rho/M$ | $\mathrm{mol \cdot cm^{-3}}$ |
| $\eta_{fM}$ | Metallic filling rate (per unit volume), defined as $\eta_{fM} = N_{\mathrm{avo}} \rho_{\mathrm{mol}} \left(\frac{4}{3} \pi r_M^3\right)$ where $N_{\mathrm{avo}}$ is the Avogadro's constant | - |
| $B$ | Bulk modulus | GPa |
| $G$ | Shear modulus | GPa |
| $\nu$ | Poisson's ratio | - |
| $\kappa$ | Thermal conductivity | $\mathrm{W \cdot m^{-1} \cdot K^{-1}}$ |
| $\alpha$ | Thermal expansion coefficient (linear not volumetric) | $\mathrm{K^{-1}}$ |
| $\theta_{\mathrm{D}}$ | Debye temperature | K |
| $\chi_m$ | Molar susceptibility | $\mathrm{m^3 \cdot mol^{-1}}$ |
| $n_c$ | Coordination number to neighbor metal atoms within 3.5 Å for each crystallographic site ($A$, $B$, and $C$). For example, in the compound LaMgNi$_4$, $A$, $B$, and $C$ correspond to La, Mg, and Ni, respectively. | - |
| $\Omega$ | Solid angle per face constructed in polyhedra of $XY_n$ ($X, Y = A, B$, and $C$ crystallographic sites; $X$-centered; and $X - Y$ distance shorter than 3.5 Å). Estimated from the parent structures listed in **Figure 2b**. | radian |
| $\Omega_\sigma$ | Standard deviation across $\Omega$ for each crystallographic site ($A, B$, and $C$). Estimated from the parent structures listed in **Figure 2b**. | radian |
| $V_p$ | Volume of polyhedra of $XY_n$ for each crystallographic site ($A, B$, and $C$). Estimated from the parent structures listed in **Figure 2b**. | Å$^3$ |
| $\langle \cdots \rangle$ | Average over the constituent metal ions | Same with the unit of $\cdots$ |
| $\sigma(\cdots)$ | Standard deviation over the constituent metal ions | Same with the unit of $\cdots$ |
| $r(\cdots)$ | Skewness over the constituent metal ions | - |
| $R_{\text{Tr}}$ | Compositional fraction of transition metal elements | - |
| $R_{\text{Tr(IV)}}$ | Compositional fraction of the fourth-row transition metal elements (Sc, …, Zn) | - |
| $R_{\text{Tr+RE}}$ | Compositional fraction of transition and rare-earth metal elements (La, …, Lu) | - |

**Key Descriptors Governing $w$ and $P_{\text{eq,RT}}$**

Given the strong predictive performance of white-box modeling approach we adopted (see the section "**Benchmarking of Regression Models**" in the **Supplementary Information**),[19] further symbolic regression models were reconstructed for the target metrics $\log_{10}(w/\langle M \rangle)$ and $\log_{10} P_{\text{eq,RT}}$, using the full dataset without reserving a separate test set. **Figure 3a** demonstrates the strong predictive performance of the model for $\log_{10}(w/\langle M \rangle)$, achieving $R^2 = 0.754$, RMSE $= 0.151$ $\log_{10}[\% \cdot \text{mol} \cdot \text{g}^{-1}]$, and MAE $= 0.0868$ $\log_{10}[\% \cdot \text{mol} \cdot \text{g}^{-1}]$ over the entire dataset ($n_{\text{data}} = 706$). **Figures 3b and 3c** present the partial dependence plots (PDPs) for two most important descriptors for $w$. In each case, the PDP curve is obtained by fixing all other features to their average values. As a result, the experimental and regressed data points, which reflect the full compositional effects and multicollinearity among features (see the section **"Pearson Correlation Heatmap"** in the **Supplementary Information**), do not necessarily follow the PDP curve. Importantly, PDPs isolate the effect of each descriptor by fixing all others, thereby revealing intrinsic relationships beyond correlations in the dataset. Nevertheless, comparing the PDP trends with the corresponding data points provides insight into the underlying physics governing the target metrics.

In **Figure 3b**, the PDP alone suggests that smaller $\langle r_M \rangle$, which is often associated with lower atomic mass favors higher $w$. However, in real materials, excessively small $\langle r_M \rangle$ can instead induce steric constraints, requiring lattice expansion to accommodate interstitial hydrogens. As a result, an optimal value emerges around $\langle r_M \rangle \sim 1.47$ Å, whereas larger values lead to expanded interstitial sites that are unable to effectively stabilize hydrogen, likely due to geometric mismatch between the interstitial cage size and the effective size required for hydrogen accommodation. Notably, in the BCC structure, $\langle r_M \rangle = 1.47$ Å yields a geometric maximum hard-sphere radius of 0.43 Å for interstitial hydrogen in tetrahedral sites. In **Figure 3c**, the PDP suggests that lower thermal conductivity $\langle \kappa \rangle$, which is linked to electronic structure near the Fermi level and the strength of metallic bonding, and typically associated with softer lattice structures, correlates with higher $w$. Here, $\langle \kappa \rangle$ is interpreted not as an isolated causal factor, but as an effective descriptor reflecting coupled electronic-structure and lattice-related characteristics of the host metal framework. In practice, however, $\langle \kappa \rangle$ exhibits a more nuanced influence. In contrast, for $\langle \kappa \rangle >$ 110 W · m$^{-1}$ · K$^{-1}$, $w$ shows little variation, indicating that the effect of $\langle \kappa \rangle$ becomes saturated in this regime. Taken together, these results indicate that **$w$ is maximized when $\langle r_M \rangle$ is tuned to approximately $1.47$ Å in conjunction with relatively low $\langle \kappa \rangle$.**

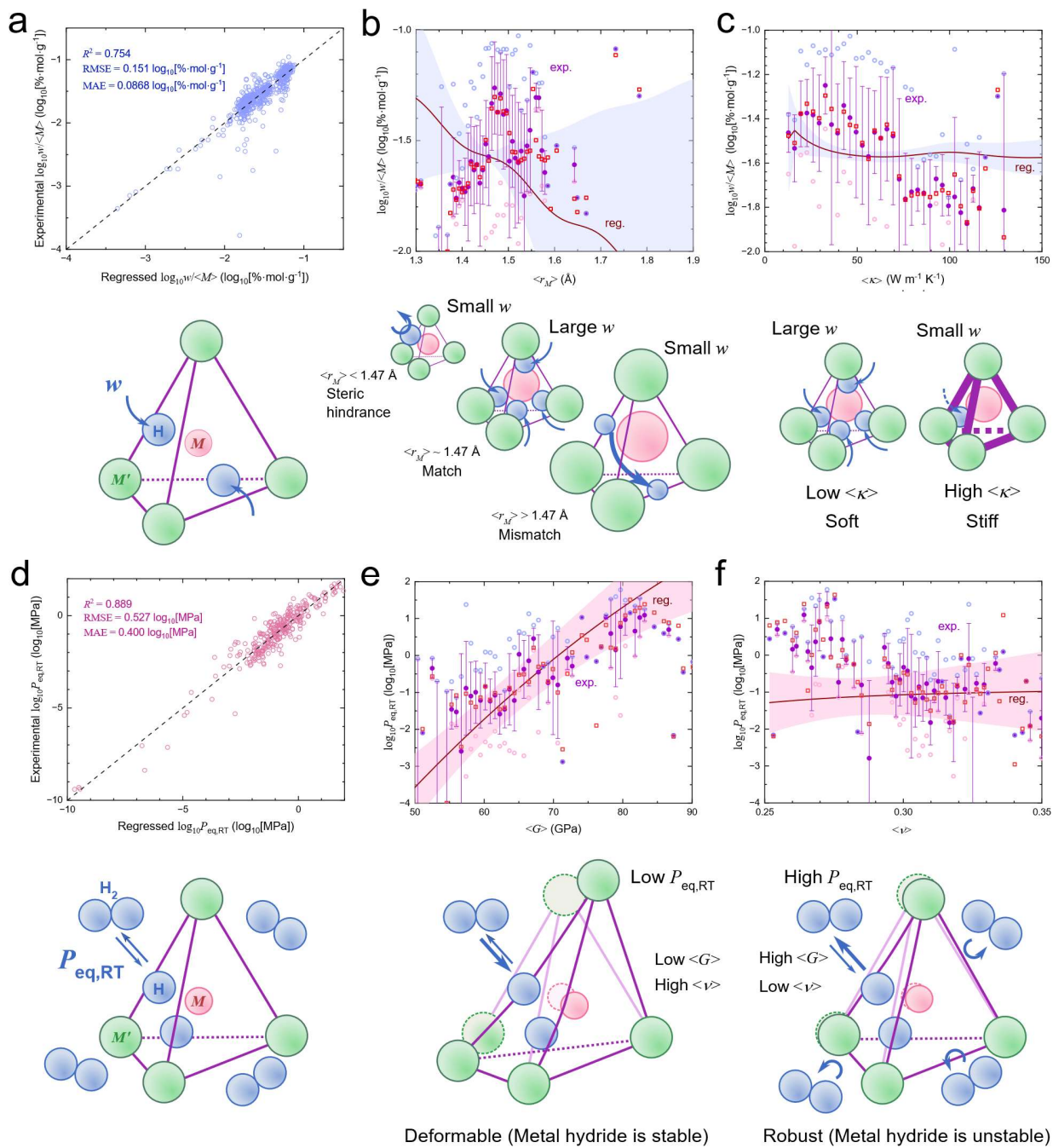

**Figure 3** Predictive performance and descriptor analysis for hydrogen capacity and equilibrium pressure. (a) Parity plot for $\log_{10}(w/\langle M \rangle)$ using the symbolic regression model. (b) and (c) Partial dependence plots (PDPs) for the two most important descriptors governing $\log_{10}(w/\langle M \rangle)$. (d) Parity plot for $\log_{10} P_{\text{eq,RT}}$. (e) and (f) PDPs for key descriptors governing $\log_{10} P_{\text{eq,RT}}$. In the PDPs of (b), (c), (e), and (f), the regression curve ("reg.") is obtained by fixing all other features to their average values. Experimental data ("exp.") are partitioned into 100 bins along the horizontal axis, where markers denote the minimum (represented by a magenta circle), average (a purple circle; with standard deviation represented by a purple bar), and maximum values (a cyan circle) within each bin, as well as the averaged regressed value (a red rectangular). The shaded region represents the variation across individual symbolic models within the ensemble. Schematic illustrations are presented below each panel to define $w$ and $P_{\text{eq,RT}}$ and to illustrate the relationships between the descriptors and the corresponding properties.

**Figure 3d** demonstrates the strong predictive performance of the model for $\log_{10} P_{\text{eq,RT}}$, achieving $R^2 = 0.889$, RMSE $= 0.527\ \log_{10}[\text{MPa}]$, and MAE $= 0.400\ \log_{10}[\text{MPa}]$ over the entire dataset ($n_{\text{data}} = 299$). **Figures 3e and 3f** present the PDPs for two most important descriptors for $\log_{10} P_{\text{eq,RT}}$. In **Figure 3e**, both the PDP and experimental and regressed data points show good agreement. A higher $\langle G \rangle$ indicates a more rigid lattice, which increases the elastic energy penalty associated with interstitial hydrogen insertion and thereby destabilizes hydride formation and leading to higher $P_{\text{eq,RT}}$. In **Figure 3f**, the PDP suggests that the influence of $\langle v \rangle$ on $P_{\text{eq,RT}}$ is relatively weak. In real materials, however, $P_{\text{eq,RT}}$ decreases rapidly when $\langle v \rangle$ is below approximately 0.3 and becomes nearly constant above this threshold. This behavior suggests that materials with higher $\langle v \rangle$ are more mechanically compliant and can more readily accommodate lattice deformation during hydride formation, consistent with the negative correlation between $\langle G \rangle$ and $\langle v \rangle$ shown in the section **"Pearson Correlation Heatmap"** in the **Supplementary Information**. Taken together, **these results indicate that lattice rigidity plays a central role in governing equilibrium pressure, with stiffer crystal structures leading to higher $P_{\text{eq,RT}}$.**

**Materials Design for High $w$ at Practical $P_{\text{eq,RT}}$**

All *DigHyd* entries with multi-temperature PCT data ($n_{\text{data}} = 299$) were subjected to optimization toward higher $w$ under the target condition of $P_{\text{eq,RT}} \sim 0.1$ MPa, guided by the symbolic models. For each parent structure class, the composition with the highest predicted $w$ within the window $-1.5\ \log_{10}[\text{MPa}] < \log_{10} P_{\text{eq,RT}} < -0.5\ \log_{10}[\text{MPa}]$ was selected, resulting in 8 representative cases. **Figures 4a-4h** are arranged in descending order of the optimized $w$, corresponding to the following structure types: BCC > Laves (C14) > LaMgNi$_4$ > La$_2$MgNi$_9$ (PuNi$_3$) > TiFe > LaNi$_5$ > Laves (C15) > Ce$_2$Ni$_7$.

As an illustrative example, **Figure 4a** shows the optimization of the BCC-type alloy TiVNbCrNi$_2$, initially characterized by a descriptor-performance vector $\boldsymbol{p} = (\langle r_M \rangle/\text{Å}, \langle \kappa \rangle/(\text{W} \cdot \text{m}^{-1} \cdot \text{K}^{-1}), w/\%) = (1.42, 63.8, 3.5)$.[23] The optimization proceeds through a multi-step compositional pathway: first, 90% of V is replaced with Ni to yield TiVNbCrV$_{1.8}$Ni$_{0.2}$; next, 10% of Cr is substituted with Tm, resulting in TiVNbCr$_{0.9}$Tm$_{0.1}$V$_{1.8}$Ni$_{0.2}$; finally, 10% of Ti is replaced with Nb, leading to the composition Ti$_{0.9}$Nb$_{0.1}$VNbCr$_{0.9}$Tm$_{0.1}$V$_{1.8}$Ni$_{0.2}$. This sequence yields the optimized composition Ti$_{0.9}$Nb$_{1.1}$V$_{2.8}$Cr$_{0.9}$Tm$_{0.1}$Ni$_{0.2}$. with final $p = (1.49, 45.1, 7.57)$. Similar symbolic-model-guided optimization systematically shifts compositions toward enhanced $w$ across other structure types[24-30] by modulating $\langle r_M \rangle$ and $\langle \kappa \rangle$ (**Figures 4b-4h**). Here, all descriptors are updated self-consistently with composition and are not fixed to their average values, in contrast to the PDP analysis in **Figure 3**.

Across all optimized cases, two consistent design principles emerge in most cases, represented by red rectangulars in **Figures 4a-4h**. First, $\langle r_M \rangle$ **converges toward an optimal value of approximately** $1.47$ **Å, represented by dashed magenta lines. Second,** $\langle \kappa \rangle$ **either decreases or remains nearly unchanged during optimization.** These trends directly reflect the descriptor-property relationships identified in **Figures 3b and 3c**, demonstrating that the symbolic models capture transferable design rules for achieving high $w$ under practical $P_{\text{eq,RT}}$ conditions. Although the predicted values of $w$ themselves may be subject to uncertainty due to their extrapolative nature, **the proposed optimization pathways provide physically grounded and clear design directions, based on the white-box symbolic models. Also, this design pathway can be further enhanced through integration with a closed-loop discovery framework,**[31, 32] **in which symbolic models are iteratively refined using synthesis and measurement data guided by the optimization trajectories, thereby enabling more accurate, real-world-relevant predictions.**

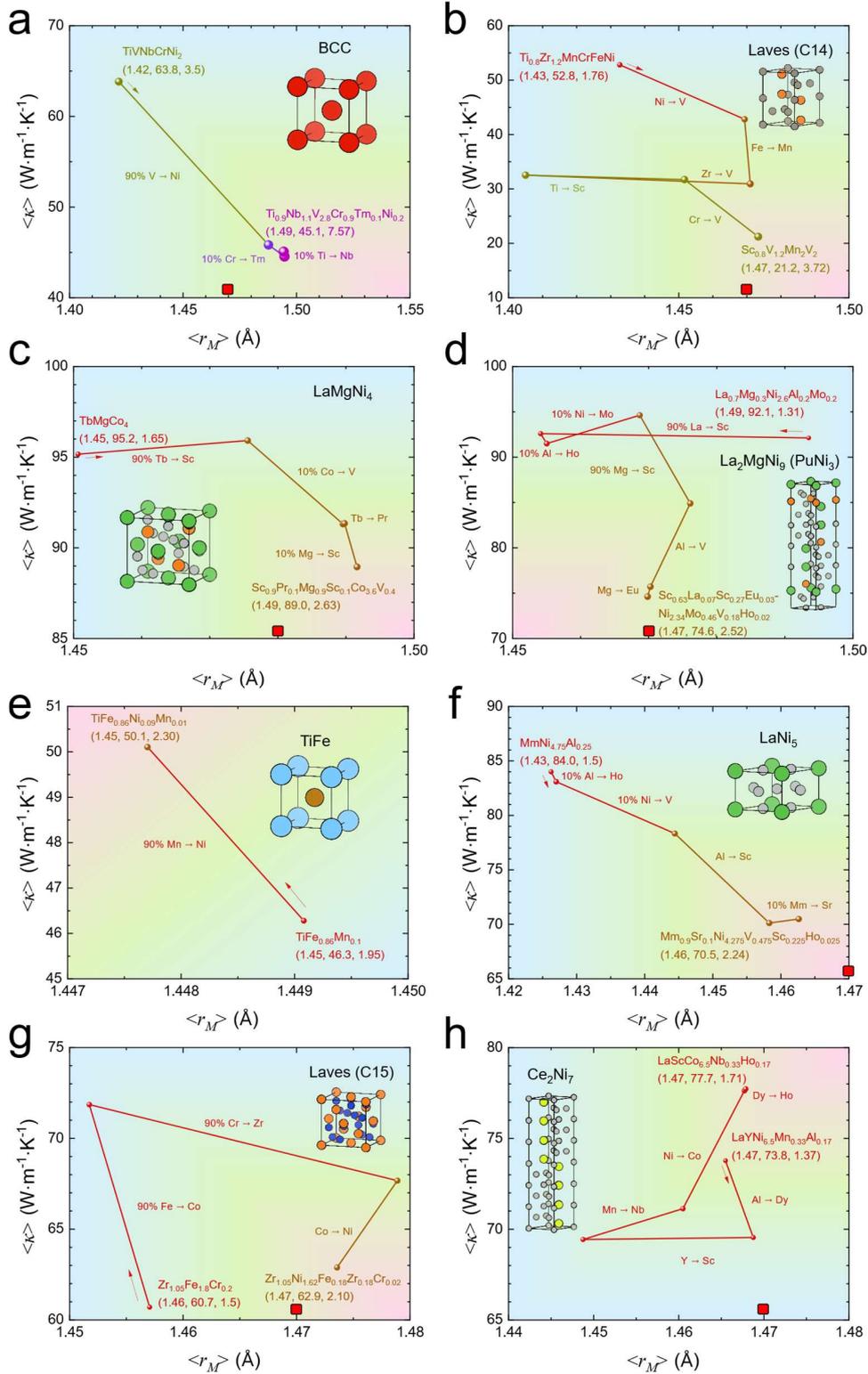

**Figure 4** Materials design pathways toward high hydrogen capacity under practical equilibrium pressure, mapped onto the ⟨κ⟩-⟨$r_M$⟩ plot. Optimization trajectories for representative parent

structure classes: (a) BCC,[23] (b) Laves (C14),[24] (c) LaMgNi$_4$,[25] (d) La$_2$MgNi$_9$ (PuNi$_3$),[26] (e) TiFe,[27] (f) LaNi$_5$,[28] (g) Laves (C15),[29] and (h) Ce$_2$Ni$_7$.[30] Each panel shows the compositional optimization pathway guided by the symbolic models within the target window $-1.5 \log_{10}[\text{MPa}] < \log_{10} P_{\text{eq,RT}} < -0.5 \log_{10}[\text{MPa}]$. The marker size and color, as well as the line color at each step, are scaled by $w$. Initial and optimized compositions are represented by descriptor-performance vectors $\boldsymbol{p} = (\langle r_M \rangle/\text{Å}, \langle \kappa \rangle/(\text{W} \cdot \text{m}^{-1} \cdot \text{K}^{-1}), w/\%)$, together with intermediate substitution steps. The red rectangular in each panel indicates a shared design target identified by symbolic models, where $\langle r_M \rangle$ converges to approximately 1.47 Å with reduced $\langle \kappa \rangle$. In most cases, $\langle \kappa \rangle$ either decreases or remains nearly constant, consistent with adherence to the design rule of $\langle r_M \rangle \to \sim 1.47$ Å. The parent crystal structures are also visually represented. The color gradient is used as a visual guide to indicate the overall direction of materials design trends, rather than representing a quantitative colormap.

**Discussion**

The present study reveals that hydrogen storage behavior in interstitial hydrides can be described by a small number of physically interpretable descriptors that govern distinct aspects of performance. In particular, $w$ is governed by the interplay between geometric and electronic-lattice factors, captured by $\langle r_M \rangle$ and $\langle \kappa \rangle$. The existence of an optimal $\langle r_M \rangle \sim 1.47$ Å reflects a geometric constraint on hydrogen accommodation, while lower $\langle \kappa \rangle$, associated with softer lattices, facilitates hydrogen incorporation. **These results indicate that maximizing $w$ requires simultaneous optimization of both interstitial geometry and lattice adaptability.** In contrast, $P_{\text{eq,RT}}$ is primarily governed by elastic properties, described by $\langle G \rangle$ and $\langle \nu \rangle$. A higher $\langle G \rangle$ increases the elastic penalty for hydrogen insertion, leading to higher equilibrium pressure, whereas larger $\langle \nu \rangle$ reflects greater mechanical compliance and lowers $P_{\text{eq,RT}}$. **Thus, $P_{\text{eq,RT}}$ is fundamentally controlled by the elastic response of the host lattice. These descriptors provide complementary representations of lattice rigidity and deformability rather than independent contributions.** We emphasize that these relationships are consistent with established physical understanding, and the present analysis provides a quantitative framework to support and organize these insights.

Taken together, these results establish a clear separation of roles between descriptors governing capacity and thermodynamics. While $w$ is controlled by geometric-electronic factors, $P_{eq,RT}$ is dictated by elastic constraints. **This separation provides a physically transparent and actionable framework for materials design under practical operating conditions.** Importantly, these descriptor relationships are consistently observed across different structure types, indicating that they define a transferable design space. **Compositions across multiple classes converge toward $\langle r_M \rangle \sim 1.47$ Å with reduced $\langle \kappa \rangle$, demonstrating the robustness of the identified design principles.**

These findings should also be understood in the broader context of descriptor discovery across hydrogen storage materials. In our previous study, we showed that across a global materials space, including both interstitial and saline hydrides, $w$ and $P_{eq,RT}$ are strongly governed by elemental electronegativity.[19] This indicates that, at the global scale, electronegativity acts as a unifying descriptor controlling hydrogen storage behavior across fundamentally different bonding types. In contrast, the present work focuses on the local materials space of interstitial hydrides, where more specific descriptors emerge, reflecting finer structural and mechanical effects. **Together, these results highlight a "hierarchical" descriptor framework, in which global trends are governed by simple chemical parameters, while local optimization requires more detailed, physically specific descriptors.**

More broadly, the present results demonstrate that symbolic regression can resolve coupled physical mechanisms into interpretable descriptor relationships, enabling a transition from empirical exploration toward mechanism-informed materials design. **This approach provides a generalizable strategy for descriptor-based materials design in hydrogen-related systems.** Furthermore, such a framework is expected to be directly applicable to saline hydrides and, more broadly, to systems beyond hydrogen storage, including hydride-based solid-state electrolytes for batteries.[33]

**Conclusion**

In this work, we established a physically interpretable, data-driven framework to uncover descriptor-property relationships governing hydrogen storage behavior in interstitial hydrides using curated PCT data and symbolic regression. **A small set of physically meaningful descriptors was identified, revealing that $w$ and $P_{eq,RT}$ are governed by fundamentally** **distinct mechanisms.** In particular, $w$ is controlled by geometric and electronic-lattice factors, whereas $P_{\text{eq,RT}}$ is dictated by elastic properties of the host lattice. **These insights were further translated into actionable materials design guidelines, enabling the identification of compositions with enhanced hydrogen capacity under practical operating conditions.** The resulting design principles are transferable across different structure types and compositional spaces, demonstrating the robustness of the descriptor-based framework. Beyond hydrogen storage, the present approach provides a general strategy for integrating interpretable machine learning with physically grounded materials design, with potential applications extending to other energy materials.

**Methods**

***DigHyd* Database**

The Digital Hydrogen Platform (*DigHyd*) is a rigorously curated database of hydrogen storage materials constructed through an AI-assisted, human-in-the-loop literature mining framework. As described in Ref. 18, *DigHyd* integrates experimentally reported data from more than 4,000 literature sources, comprising over 30,000 curated data entries on hydrogen storage properties and thermodynamic parameters. Importantly, the database spans a wide range of material classes beyond conventional metal hydrides, including interstitial hydrides, complex hydrides, ionic (saline) hydrides, multi-component and destabilized systems, as well as porous materials such as metal-organic frameworks. This breadth enables systematic analysis across fundamentally different hydrogen storage mechanisms. The curated data include gravimetric hydrogen capacity ($w$), as well as the enthalpy ($\Delta H$) and entropy ($\Delta S$) changes associated with hydrogenation reactions, primarily defined as $M + \frac{1}{2}\text{H}_2 \rightleftharpoons M\text{H}$.

**Symbolic Modeling**

While the algorithmic details of the white-box symbolic regression modeling package (*GoodRegressor*) are described in Ref. 20, a brief summary of its use in this study is provided here. For each symbolic model constructed using the *Regressor* module described in Ref. 20, the dataset was randomly split into training and validation sets with a ratio of 8: 2. Given the set of candidate descriptors listed in **Table 1**, the number of possible model combinations including interaction terms is given as $1.95 \times 10^{463}$ for the 109 default scalar transforms provided in the *Regressor* module.

Model construction was therefore performed in a staged manner. Starting from models with 20 active independent variables, the number of active variables was iteratively reduced one by one ($20 \rightarrow 19 \rightarrow 18 \rightarrow \cdots \rightarrow 1$), while progressively allowing more complex interaction terms (i.e., increasing interaction depth). At each stage, candidate models were evaluated, and the model with the highest $R^2$ over the entire dataset was selected. In this process, $\geq 12$ CPU cores were used for each symbolic model construction.

All models were generated under the so-called *full Fisher condition*, whereby the $p$-values of both the overall model ($F$-test) and all individual coefficients ($t$-tests) were required to be less than $0.05$. For each target metric ($\log_{10}(w/\langle M \rangle)$ and $\log_{10} P_{\text{eq,RT}}$), 10 symbolic models were constructed and subsequently combined into a single "stacking-ensembled" model. Key descriptors were then identified by examining features that consistently appeared across the 10 symbolic models, as well as by evaluating the magnitude of their contributions through $z$-scored coefficients. In particular, descriptors associated with terms exhibiting large average $z$-scored coefficient magnitudes across the ensemble were considered to play dominant roles in determining the target properties.

To ensure model reliability, the standard deviation across the regressed values in the ensemble was constrained to be below $0.5$ $\log_{10}[\% \cdot \text{mol} \cdot \text{g}^{-1}]$ and $0.7$ $\log_{10}[\text{MPa}]$ for $\log_{10}(w/\langle M \rangle)$ and $\log_{10} P_{\text{eq,RT}}$, respectively. Predictions exceeding these thresholds were discarded. In the 5-fold benchmark tests, the discard ratios were approximately 1% and 10% for $\log_{10}(w/\langle M \rangle)$ and $\log_{10} P_{\text{eq,RT}}$, respectively. Notably, when models were constructed using the entire dataset, no data points were discarded, as all predictions satisfied the ensemble consistency criteria. Representative examples of symbolic model ensembles, their error distributions, and comparisons across different target metrics ($\log_{10} w$ and $\log_{10}(w\langle M \rangle)$) are provided in the section "**Symbolic Models: Formulation and Error Analysis**" in the **Supplementary Information**.

**Materials Design**

For materials optimization, the *Designer* module described in Ref. 20 was employed, using the *DigHyd* dataset as input. During the search for optimization pathways from existing experimental compositions, only candidate compositions yielding ensemble standard deviations below $0.5$ $\log_{10}[\% \cdot \text{mol} \cdot \text{g}^{-1}]$ and $0.7$ $\log_{10}[\text{MPa}]$ for $\log_{10}(w/\langle M \rangle)$ and $\log_{10} P_{\text{eq,RT}}$, respectively, were considered. For the final optimized compositions, a stricter criterion was applied: only those with ensemble standard deviations below 0.1 $\log_{10}[\% \cdot \text{mol} \cdot \text{g}^{-1}]$ were retained for $\log_{10}(w/\langle M \rangle)$.

Compositional modifications were performed through controlled substitution operations while preserving the original stoichiometry. Specifically, full substitution, 90% substitution, and 10% substitution of constituent metal elements were allowed. Candidate substitution elements were required to satisfy two criteria relative to the original element: the differences in electronegativity and metallic radius must not exceed 0.5 and 0.5 Å, respectively.

**Associated content**

Supporting Information: Radial distribution function averaged over metal-alloy parent structures for hydrogen storage and the Construction of $XY_n$ Polyhedra, Pearson correlation heatmap, benchmarking of regression models, and symbolic models: formulation and error analysis

**Author information**

**Data Availability**

The data can be accessed in the Digital Hydrogen Platform (*DigHyd*: www.dighyd.org).

**Code Availability**

The source code supporting materials prediction and design in this study is openly available at https://github.com/JerryGarcia1995/OxygenIonConductor.

**Acknowledgments**

This work was supported by The Green Technologies of Excellence (GteX) Program, Japan (Grant No. JPMJGX23H1). Crystal structures were visualized using VESTA.[34]

**References**

1. Johnson, N.; Liebreich, M.; Kammen, D. M.; Ekins, P.; McKenna, R.; Staffell, I. Realistic roles for hydrogen in the future energy transition. *Nat. Rev. Clean Technol.* **2025,** *1*, 351-371. DOI: 10.1038/s44359-025-00050-4

2. Zhao, A. P.; Li, S.; Xie, D.; Wang, Y.; Li, Z.; Hu, P. J.-H.; Zhang, Q. Hydrogen as the nexus of future sustainable transport and energy system. *Nat. Rev. Electr. Eng.* **2025,** *2*, 447-446. DOI: 10.1038/s44287-025-00178-2

3. Gebretatios, A. G.; Banat, F.; Cheng, C. K. A critical review of hydrogen storage: Toward the nanoconfinement of complex hydrides from the synthesis and characterization perspectives. *Sustain. Energ. Fuels.* **2024,** *8* (22), 5091-5130. DOI: 10.1039/d4se00353e

4. Züttel, A. Materials for hydrogen storage. *Mater. Today* **2003**, *6* (9), 24-33. DOI: 10.1016/S1369-7021(03)00922-2

5. Bellosta von Colbe, J.; Ares, J.-R.; Barale, J.; Baricco, M.; Buckley, C.; Capurso, G.; Gallandat, N.; Grant, D. M.; Guzik, M. N.; Jacob, I.; Jensen, E. H.; Jensen, T.; Jepsen, J.; Klassen, T.; Lototskyy, M. V.; Manickam, K.; Montone, A.; Puszkiel, J.; Sartori, S.; Sheppard, D. A.; Stuart, A.; Walker, G.; Webb, C. J.; Yang, H.; Yartys, V.; Züttel, A.; Dornheim, M. Application of hydrides in hydrogen storage and compression: Achievements, outlook and perspectives. *Int. J. Hydrogen Energy* **2019,** *44* (15), 7780-7808. DOI: 10.1016/j.ijhydene.2019.01.104

6. Hirscher, M.; Yartys, V. A.; Baricco, M.; Bellosta von Colbe, J.; Blanchard, D.; Bowman, R. C.; Broom, D. P.; Buckley, C. E.; Chang, F.; Chen, P.; Cho, Y. W.; Crivello, J.-C.; Cuevas, F.; David, W. I. F.; de Jongh, P. E.; Denys, R. V.; Dornheim, M.; Felderhoff, M.; Filinchuk, Y.; Froudakis, G. E. Materials for hydrogen-based energy storage - Past, recent progress and future outlook. *J. Alloys Compd.* **2020**, *827*, 153548. DOI: 10.1016/j.jallcom.2019.153548

7. Wicke, E.; Brodowsky, H.; Züchner, H. Hydrogen in palladium and palladium alloys. *Hydrogen in Metals II. Topics in Applied Physics, vol 29;* Alefeld, G.; Völkl, J., Eds.; Springer, 1978; pp 73-155. DOI: 10.1007/3-540-08883-0_19

8. Sandrock, G. A panoramic overview of hydrogen storage alloys from a gas reaction point of view. *J. Alloys Compd.* **1999,** *293-295,* 877-888. DOI: 10.1016/S0925-8388(99)00384-9

9. Bowman Jr, R. C.; Fultz, B. Metallic hydrides I: Hydrogen storage and other gas-phase applications. *MRS Bulletin* **2002,** 27, 688-693. DOI: 10.1557/mrs2002.223

10. Orimo, S.; Nakamori, Y.; Eliseo, J. R.; Züttel, A.; Jensen, C. M. Complex hydrides for hydrogen storage. *Chem. Rev.* **2007,** 107, 4111-4132. DOI: 10.1021/cr0501846

11. Sakintuna, B.; Lamari-Darkrim, F.; Hirscher, M. Metal hydride materials for solid hydrogen storage: A review. *Int. J. Hydrogen Energy* **2007,** *32* (9), 1121-1140. DOI: 10.1016/j.ijhydene.2006.11.022

12. Jain, I. P.; Jain, P.; Jain, A. Novel hydrogen storage materials: A review of lightweight complex hydrides. *J. Alloys Compd.* **2010,** *503* (2), 303-339. DOI: 10.1016/j.jallcom.2010.04.250

13. Jain, I. P.; Lal, C.; Jain, A. Hydrogen storage in Mg: A most promising material. *Int. J. Hydrogen Energy* **2010,** *35* (10), 5133-5144. DOI: 10.1016/j.ijhydene.2009.08.088

14. Scarpati, G.; Frasci, E.; Di Ilio, G.; Jannelli, E. A comprehensive review on metal hydrides-based hydrogen storage systems for mobile applications. *J. Energy Storage* **2024,** *102*, 113934. DOI: 10.1016/j.est.2024.113934

15. Nemukula, E.; Mtshali, C. B.; Nemangwele, F. Metal hydrides for sustainable hydrogen storage: A review. *Int. J. Energy Res.* **2025**, 6302225. DOI: 10.1155/er/6300225

16. Gao, Z.; Yang, X.; Zhuang, Z.; Zhang, Y.; Cai, J.; Li, Y.; Fu, W.; Li, H.; Yang, W. Catalytic strategies and mechanisms for enhancing $MgH_2$ solid-state hydrogen storage. *Chem Catal.* **2026,** *6*, 101692. DOI: 10.1016/j.checat.2026.101692

17. Zhang, D.; Jia, X.; Tran, H. B.; Jang, S. H.; Zhang, L.; Sato, R.; Hashimoto, Y.; Sato, T.; Konno, K.; Orimo, S.; Li, H. "DIVE" into hydrogen storage materials discovery with AI agents. *Chem. Sci.* **2026,** 17, 3031-3042. DOI: 10.1039/D5SC09921H

18. Jang, S.-H.; Zhang, D.; Jia, X.; Tran, H. B.; Zhang, L.; Sato, R.; Hashimoto, Y.; Sato, T.; Konno, K.; Orimo, S.; Li, H. Digital hydrogen platform (DigHyd): A rigorously curated database for hydrogen storage materials empowered by AI-assisted literature mining. *arXiv*, March 14, 2026. DOI: 10.48550/arXiv.2603.14139

19. Jang, S.-H.; Zhang, D.; Tran, H. B.; Jia, X.; Konno, K.; Sato, R.; Orimo, S.; Li, H. Physically interpretable descriptors drive the materials design of metal hydrides for hydrogen storage. *Chem. Sci.* **2025,** 16, 23111-23120. DOI: 10.1039/D5SC07296D

20. Jang, S.-H. GoodRegressor: A hierarchical inductive bias for navigating high-dimensional compositional space. *arXiv*, February 20, 2026. DOI: 10.48550/arXiv.2510.18325

21. World Health Organization & International Programme on Chemical Safety. *Beryllium: Health and safety guide*; World Health Organization, 1990. https://iris.who.int/handle/10665/40004

22. Akiba, E.; Okada, M. Metallic hydrides III: Body-centered-cubic solid-solution alloys. *MRS Bull.* **2002,** *27* (9), 699-703. DOI: 10.1557/mrs2002.225

23. Cheng, B.; Kong, L.; Cai, H.; Li, Y.; Zhao, Y.; Wan, D.; Xue, Y. Exploring microstructure variations and hydrogen storage characteristics in TiVNbCrNi high-entropy alloys with different Ni incorporation. *Int. J. Hydrog. Energy* **2024,** *72*, 29-40. DOI: 10.1016/j.ijhydene.2024.05.317

24. Enblom, V.; Clulow, R.; Ha, T.-J.; Witman, M. D.; Way, L. E.; Han, S. J.; Brant Carvalho, P. H. B.; Stavila, V.; Suh, J.-Y.; Sahlberg, M.; Fadonougbo, J. O. A combined experimental and machine learning exploration of $Ti_{2-x}Zr_xMnCrFeNi$ high entropy Laves hydrides. Mater., 2025, 40, 102414. DOI: 10.1016/j.mtla.2025.102414

25. Shtender, V. V.; Paul-Boncour, V.; Denys, R. V.; Crivello, J.-C.; Zavaliy, I. Y. $TbMgNi_{4-x}Co_x$-$(H,D)_2$ System. I: Synthesis, hydrogenation properties, and crystal and electronic structures. *J. Phys. Chem. C* **2020,** *124* (1), 196-204. DOI: 10.1021/acs.jpcc.9b10252

26. Zhang, X. B.; Sun, D. Z.; Yin, W. Y.; Chai, Y. J.; Zhao, M. S. Crystallographic and electrochemical characteristics of $La_{0.7}Mg_{0.3}Ni_{3-x}(Al_{0.5}Mo_{0.5})_x$ ($x$ = 0-0.4) hydrogen storage alloys. *Electrochim. Acta.* **2005,** *50* (16-17), 3407-3413. DOI: 10.1016/j.electacta.2004.12.020

27. Qu, H.; Du, J.; Pu, C.; Niu, Y.; Huang, T.; Li, Z.; Lou, Y.; Wu, Z. Effects of Co introduction on hydrogen storage properties of Ti-Fe-Mn alloys. *Int. J. Hydrog. Energy* **2015,** *40* (6), 2729-2735. DOI: 10.1016/j.ijhydene.2014.12.089

28. Molinas, B.; Pontarollo, A.; Scapin, M.; Peretti, H.; Melnichuk, M.; Corso, H.; Aurora, A.; Gatttia, D. M.; Montone, A. The optimization of $MmNi_{5-x}Al_x$ hydrogen storage alloy for sea or lagoon navigation and transportation. *Int. J. Hydrog. Energy* **2016,** *41* (32), 14484-14490. DOI: 10.1016/j.ijhydene.2016.05.222

29. Zhou, C.; Wang, H.; Ouyang, L. Z.; Liu, J. W.; Zhu, M. Achieving high equilibrium pressure and low hysteresis of Zr-Fe based hydrogen storage alloy by Cr/V substitution. *J. Alloys Compd.* **2019,** *806*, 1436-1444. DOI: 10.1016/j.jallcom.2019.07.170

30. Jensen, E. H.; Lombardo, L.; Girella, A.; Guzik, M. N.; Züttel, A.; Milanese, C.; Whitfield, P.; Noréus, D.; Satori, S. The effect of Y content on structural and sorption properties of $A_2B_7$-type phase in the La-Y-Ni-Al-Mn system. *Molecules* **2023,** *28* (9), 3749. DOI: 10.3390/molecules28093749

31. Zhang, D.; Jia, X.; Wang, Y.; Liu, H.; Wang, Q.; Jang, S.-H.; Shah, D.; Ye, S.; Tran, H. B.; Li, H. Digital materials ecosystem: from databases to AI agents for autonomous discovery. *Chem. Sci.* **2026,** *17*, 5782-5804. DOI: 10.1039/D5SC09229A

32. Zhang, D.; Chen, Y.; Liu, C.; Liu, Y.; Xin, H.; Peng, J.; Ou, P.; Li, H. Accelerating catalyst materials discovery with large artificial intelligence models. *Angew. Chem. Int. Ed.* **2026,** e26150. DOI: 10.1002/anie.202526150

33. Wang, Q.; Yang, F.; Wang, Y.; Zhang, D.; Sato, R.; Zhang, L.; Cheng, J.; Yan, Y.; Chen, Y.; Kisu, K.; Orimo, S.; Li, H. Unraveling the complexity of divalent hydride electrolytes in solid-state batteries via a data-driven framework with large language model. *Angew. Chem. Int. Ed.* **2025,** *64* (5), e202506573. DOI: 10.1002/anie.202506573

34. Momma, K.; Izumi, F. VESTA 3 for three-dimensional visualization of crystal, volumetric and morphology data. *J. Appl. Cryst.* **2011,** 44, 1272-1276. DOI: 10.1107/S0021889811038970

**ToC**

White-box "symbolic" regression modeling

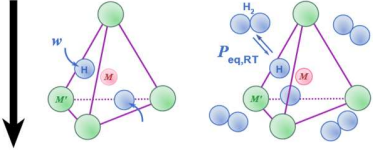

Detect "important" descriptors

$\langle r_M \rangle \quad \langle \kappa \rangle \quad \langle G \rangle \quad \langle v \rangle$

Physically-grounded materials design

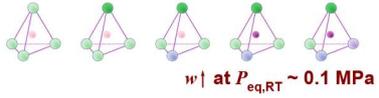

$w \uparrow$ at $P_{eq,RT} \sim 0.1$ MPa

Supplementary Information

# A unified descriptor framework for hydrogen storage capacity and equilibrium pressure in interstitial hydrides

*Seong-Hoon Jang*[*,1, 2], *Di Zhang*[1,3], *Xue Jia*[1], *Hung Ba Tran*[1], *Linda Zhang*[1,3], *Ryuhei Sato*[4], *Yusuke Hashimoto*[3], *Yusuke Ohashi*[5], *Toyoto Sato*[5], *Kiyoe Konno*[6], *Shin-ichi Orimo*[*1, 5], *and Hao Li*[*1]

[1] Advanced Institute for Materials Research (WPI-AIMR), Tohoku University, Sendai 980-8577, Japan

[2] Unprecedented-scale Data Analytics Center, Tohoku University, Sendai 980-8578, Japan

[3] Frontier Research Institute for Interdisciplinary Sciences (FRIS), Tohoku University, Sendai 980-8577, Japan

[4] Department of Materials Engineering, The University of Tokyo, Tokyo 113-8656, Japan

[5] Institute for Materials Research, Tohoku University, Sendai, 980-8577

[6] Institute of Fluid Science, Tohoku University, Sendai, 980-8577, Japan

Contents:

- Radial Distribution Function Averaged over Metal-alloy Parent Structures for Hydrogen Storage and the Construction of $XY_n$ Polyhedra

- Pearson Correlation Heatmap

- Benchmarking of Regression Models

  Details of Conventional Machine Learning Approaches

- Symbolic Models: Formulation and Error Analysis

  Symbolic Model for $\log_{10}(w/\langle M \rangle)$

  Symbolic Model for $\log_{10} P_{\text{eq,RT}}$

  Stacking-ensembled Symbolic Models for $\log_{10} w$ and $\log_{10}(w\langle M \rangle)$

  Error Distributions of Stacking-ensembled Models for $\log_{10}(w/\langle M \rangle)$ and $\log_{10} P_{\text{eq,RT}}$

**Radial Distribution Function Averaged over Metal-alloy Parent Structures for Hydrogen Storage and the Construction of $XY_n$ Polyhedra**

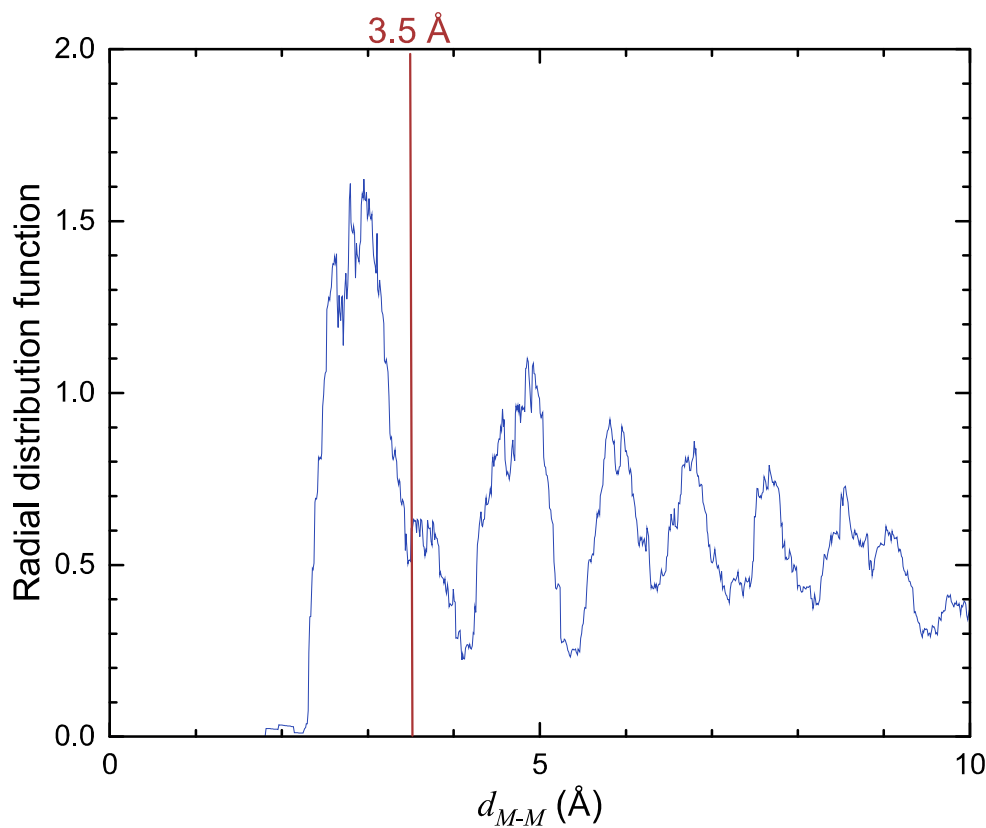

**Supplementary Figure 1.** Averaged radial distribution functions of metal-alloy parent structures. Radial distribution function averaged over representative metal-alloy parent structures used in this study. The vertical red line at 3.5 Å indicates the cutoff distance adopted for defining neighboring metal atoms in the construction of coordination polyhedra.

In **Supplementary Figure 1,** The first major peak corresponds to nearest-neighbor metallic bonding, while the shoulder region around 3.5 Å represents the upper bound of local coordination environments across different structures. Based on this observation, a cutoff distance of 3.5 Å was selected to consistently define neighboring atoms for structural descriptor construction. This choice ensures that the coordination polyhedra capture the relevant local environments while avoiding inclusion of more distant, weakly interacting atoms.

For the construction of polyhedra $XY_n$ (where $X$ and $Y$ are metal atoms, with $X$ at the center), all $Y$ atoms within a distance of 3.5 Å from $X$ were selected. Three-dimensional convex hulls were then generated. To achieve this, all possible triangular faces $\left(\binom{n}{3}\right)$ were enumerated and subsequently evaluated to determine whether they belonged to the outermost surface or the interior of the polyhedron, based on the orientation of their perpendicular vectors. Subsequently, the included angles between all adjacent faces of the convex hull were examined. If two adjacent faces were found to be coplanar (i.e., lying in the same plane), they were merged into a single unified face. This integration process was iteratively repeated until no further coplanar faces remained. This procedure ensures that the polyhedra are uniquely defined in all cases. Following construction, the validity of each polyhedron was confirmed by verifying that the sum of the solid angles of all faces equals $4\pi$.

**Pearson Correlation Heatmap**

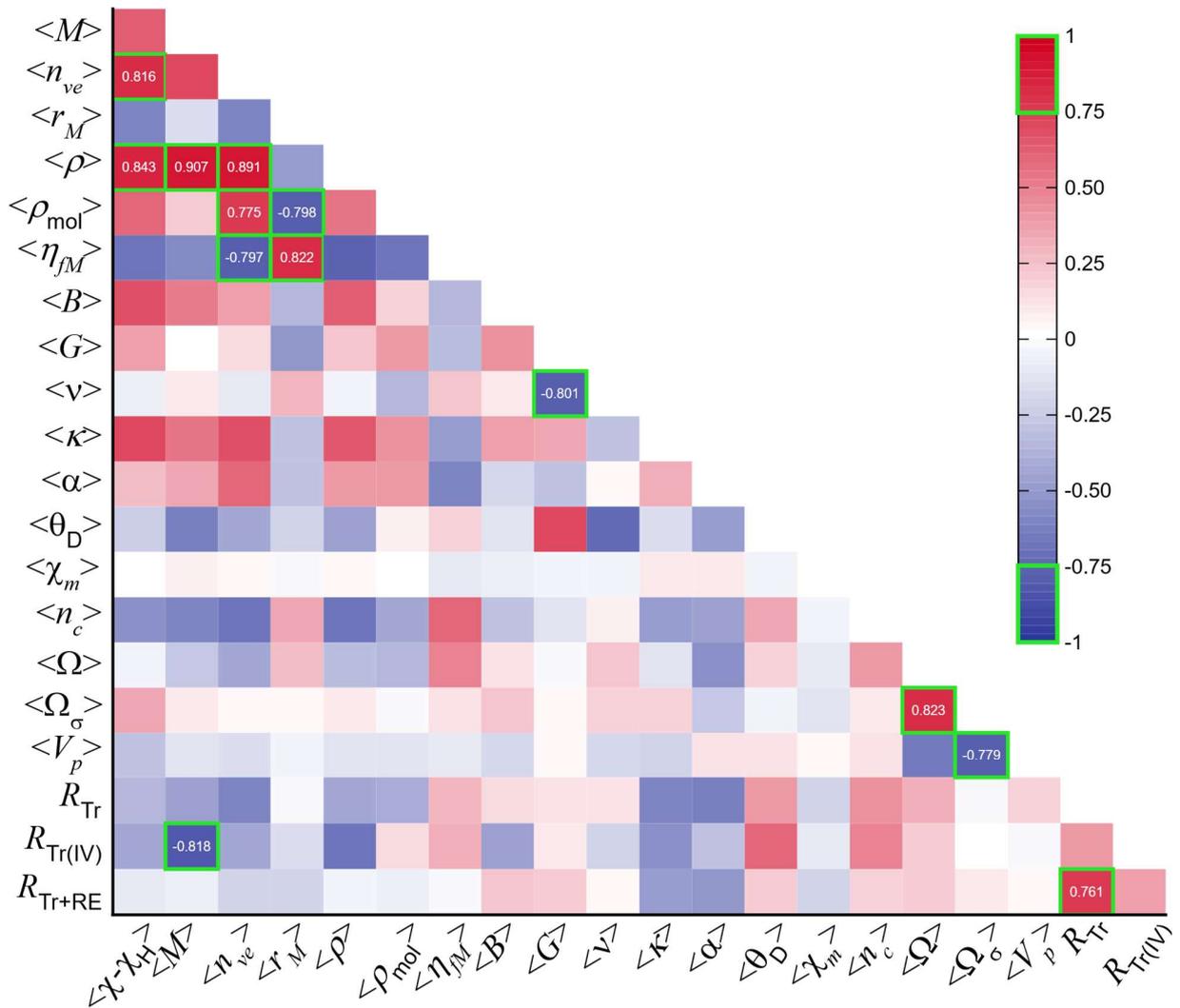

**Supplementary Figure 2** Heatmap of Pearson correlation coefficients ($r_{\text{col}}$) among the selected descriptors. Strong correlations ($|r_{\text{col}}| > 0.75$) are highlighted.

**Supplementary Figure 2** shows the heatmap of Pearson correlation coefficients ($r_{\text{col}}$) among all feature pairs. Several pairs exhibit strong correlations ($|r_{\text{col}}| > 0.75$), indicating intrinsic coupling among descriptors. For example, the physical descriptors $\langle G \rangle$ and $\langle v \rangle$ show a strong negative correlation ($r_{\text{col}} = -0.801$), reflecting the fundamental mechanical response of solids in which materials with higher $G$ tend to resist transverse deformation and therefore exhibit lower $v$. The interdependence patterns provide a physical basis for the descriptor analysis presented in the main text.

**Benchmarking of Regression Models**

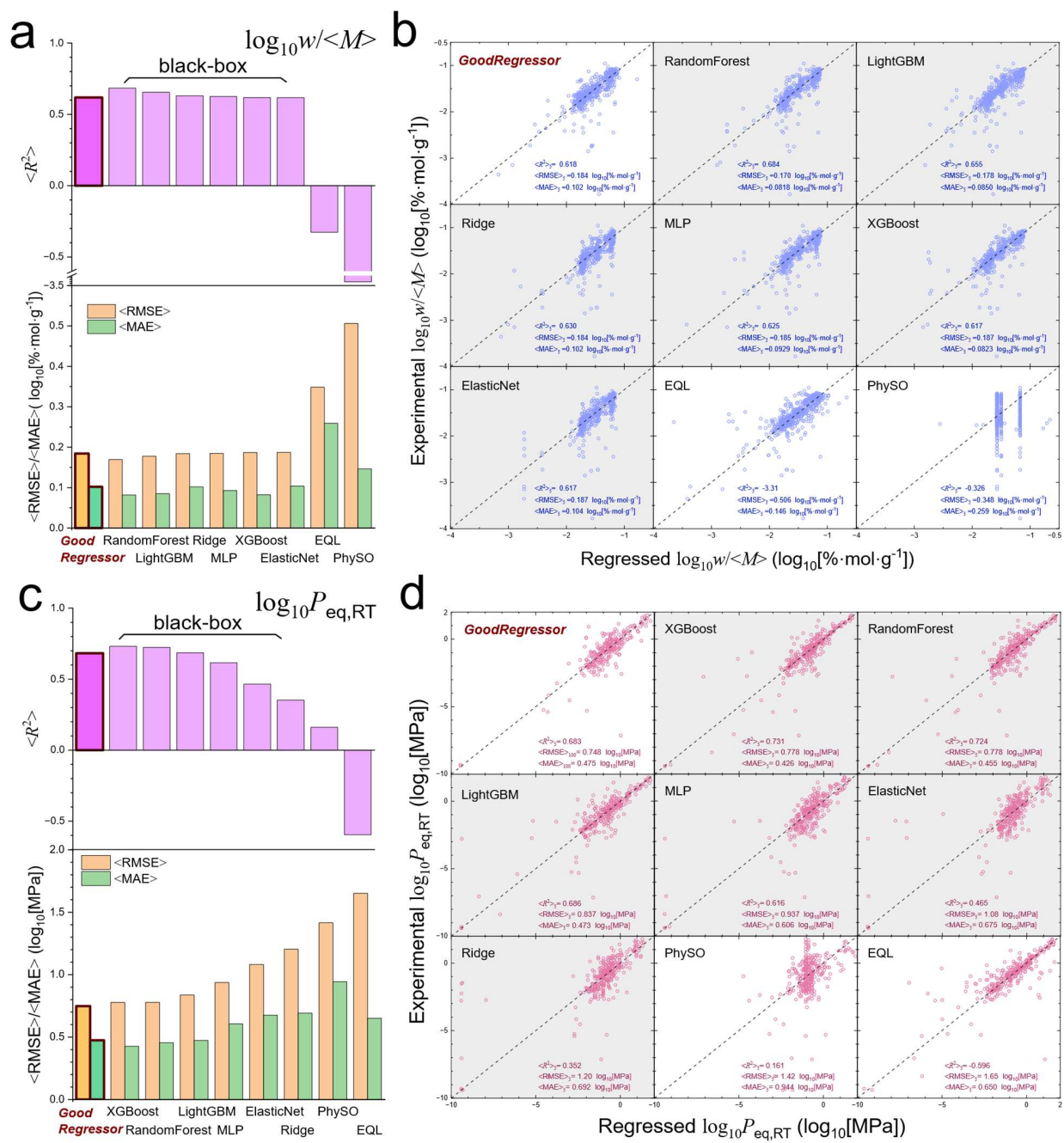

**Supplementary Figure 3** Benchmark performance of *GoodRegressor* and other machine learning models. Comparison of stacking-ensembled symbolic regression models generated by using *GoodRegressor*[S1] with conventional machine learning approaches, Ridge,[S2] ElasticNet,[S3] MLP,[S4] RandomForest,[S5] XGBoost,[S6] LightGBM,[S7] EQL,[S8] and Φ-SO,[S9] for predicting (a, b) $w$ and (c, d)

$P_{eq,RT}$. (a) and (c) show averaged benchmark metrics across five different test folds, that is, coefficient of determination ($\langle R^2 \rangle_5$), root mean square error ($\langle \text{RMSE} \rangle_5$), and mean absolute error ($\langle \text{MAE} \rangle_5$), and present schematic illustrations defining $w$ and $P_{eq,RT}$. (b) and (d) present parity plots comparing experimental and predicted values for each model. In (b) and (d), the black-box models are illustrated as gray panels for visual distinction.

To assess whether *GoodRegressor*, a white-box symbolic regression framework,[S1] can simultaneously achieve high predictive accuracy and interpretability, benchmark tests were performed for two target metrics: $\log_{10}(w/\langle M \rangle)$ and $\log_{10} P_{eq,RT}$. The comparison included black-box models, Ridge,[S2] ElasticNet,[S3] MLP,[S4] RandomForest,[S5] XGBoost,[S6] and LightGBM,[S7] as well as other white-box baselines, EQL[S8] and Phy-SO,[S9] as illustrated in **Supplementary Figure 3**. Technical details are provided in subsequent sections. All models were evaluated using 5-fold cross-validation with hyperparameter optimization, employing identical data splits to ensure fair comparison. For instance, in the *GoodRegressor* case, a fold (20%) was held out as the test set in each iteration, while the remaining data were used for training and validation. Within this subset, 10 independent random 8:2 train-validation splits were generated to construct a stacking-ensembled symbolic regression model. Repeating this procedure across all folds yielded 50 individual symbolic models and 5 ensemble models in total, corresponding to overall data usage of 64 % for training, 16 % for validation, and 20 % for testing.

In **Supplementary Figure 3a**, the benchmark results for the $\log_{10}(w/\langle M \rangle)$ dataset are presented in terms of three metrics: the averaged coefficient of determination ($\langle R^2 \rangle_5$), the averaged root mean square error ($\langle \text{RMSE} \rangle_5$), and the averaged mean absolute error ($\langle \text{MAE} \rangle_5$) across the five validation folds. **The *GoodRegressor* model achieved performance comparable to state-of-the-art black-box machine learning models while outperforming all other white-box baselines.** Specifically, it achieved $\langle R^2 \rangle_5 = 0.618$, $\langle \text{RMSE} \rangle_5 = 0.184 \log_{10}[\% \cdot \text{mol} \cdot \text{g}^{-1}]$, and $\langle \text{MAE} \rangle_5 = 0.102 \log_{10}[\% \cdot \text{mol} \cdot \text{g}^{-1}]$. Meanwhile, the other models exhibited performance ($\langle R^2 \rangle_5 \leq 0.684$, $\langle \text{RMSE} \rangle_5 \geq 0.170 \log_{10}[\% \cdot \text{mol} \cdot \text{g}^{-1}]$, and $\langle \text{MAE} \rangle_5 \geq 0.0818 \log_{10}[\% \cdot \text{mol} \cdot \text{g}^{-1}]$). The performance ranking among these was as follows: RandomForest > LightGBM > Ridge > MLP > *GoodRegressor* > XGBoost > ElasticNet > EQL > PhySO. The parity plots in **Supplementary Figure 3b** as well further confirm that the *GoodRegressor* model provides excellent predictive accuracy.

**Supplementary Figure 3c** presents the benchmark results for the dataset of $\log_{10} P_{eq,RT}$. **Again, the *GoodRegressor* model demonstrated performance comparable to black-box machine learning methods while outperforming other white-box baselines.** Specifically, it achieved $\langle R^2 \rangle_5 = 0.683$, $\langle \text{RMSE} \rangle_5 = 0.748 \log_{10}[\text{MPa}]$, and $\langle \text{MAE} \rangle_5 = 0.575 \log_{10}[\text{MPa}]$. The other models recorded comparable or lower performance ($\langle R^2 \rangle_5 \leq 0.731$, $\langle \text{RMSE} \rangle_5 \geq 0.778 \log_{10}[\text{MPa}]$, and $\langle \text{MAE} \rangle_5 \geq 0.426 \log_{10}[\text{MPa}]$). The performance ranking among these was as follows: XGBoost > RandomForest > LightGBM > *GoodRegressor* > MLP > ElasticNet > Ridge > PhySO > EQL. The parity plots in **Supplementary Figure 3d** as well further confirm that *GoodRegressor* provides excellent predictive accuracy. Taken together, *GoodRegressor* represents a unique approach that simultaneously achieves reliable predictive performance and inherent white-box interpretability.

**Details of conventional machine learning approaches.** To evaluate the predictive performance of various regression algorithms on the dataset, a standardized benchmarking pipeline was implemented in Python. This framework provides a uniform and unbiased comparison between conventional machine-learning and symbolic regression approaches. The evaluation follows a nested cross-validation design with systematic hyperparameter optimization, ensuring fair and reproducible comparison across models. Each model underwent 5-fold nested cross-validation. The details of parameters are given below.

**a. Ridge**

Core library: scikit-learn

Search parameters: $\alpha \in [10^{-4}, 10^{3}]$

Iterations: 80

**b. ElasticNet**

Core library: scikit-learn

Search parameters: $\alpha \in [10^{-4}, 10^{1}]$, $l_1$ ratio $\in [0, 1]$

Iterations: 80

**c. MLP**

Core library: scikit-learn

Search parameters: Hidden layers: (256, 128), (256, 128, 64), (512, 256, 128); activation: ReLU or tanh; $\alpha \in [10^{-6}, 10^{-2}]$; learning rate $\in [3 \times 10^{-4}, 3 \times 10^{-2}]$; max_iter = 4000

Iterations: 80

**d. RandomForest**

Core library: scikit-learn

Search parameters: n_estimators ∈ [800, 2000], max_depth ∈ [6, 28], min_samples_split ∈ [2, 12], min_samples_leaf ∈ [1, 6], max_features ∈ [0.3, 0.7]

Iterations: 100

**e. XGBoost**

Core library: xgboost

Search parameters: n_estimators ∈ [1200, 3000], learning rate ∈ [0.01, 0.2], max_depth ∈ [3, 12], subsample ∈ [0.6, 1.0], reg_lambda ∈ [1, 80], gamma ∈ [$10^{-9}$, $10^{-1}$]

Iteration: 140

**f. LightGBM**

Core library: lightgbm

Search parameters: n_estimators ∈ [1500, 4000], learning rate ∈ [0.01, 0.2], num_leaves ∈ [31, 255], min_child_samples ∈ [5, 120], feature/bagging fraction ∈ [0.6, 1.0], $\lambda_1$, $\lambda_2$ ∈ [$10^{-3}$, 10]

Iteration: 140

**g. EQL**

Epochs: 600

Learning rate: $1 \times 10^{-3}$

$L_1$ penalty: $1 \times 10^{-3}$ on output weights and projection layers

Activation functions: {sin, cos, exp, log, erf, square, cube, linear, multiplicative interactions}

Term constraint: maximum 20 active symbolic terms via adaptive top-K masking

**i. Φ-SO**

Each symbolic search ran for 60 epochs, operating on symbolic operators {mul, add, sub, div, n², sqrt, neg, exp, log, sin, cos}.

**Symbolic Models: Formulation and Error Analysis**

In this section, as representative examples, one of the ten symbolic regression models for each target metric ($\log_{10}(w/\langle M \rangle)$ and $\log_{10} P_{\text{eq,RT}}$) is provided, and error distributions for both target properties are provided as well. Additionally, "inferior" stacking-ensembled symbolic models for $\log_{10} w$ and $\log_{10}(w\langle M \rangle)$ are represented.

## a. Symbolic Model for $\log_{10}(w/\langle M \rangle)$

**Supplementary Table 1** An example of symbolic model for $\log_{10}(w/\langle M \rangle)$. Here, $\chi_{\Delta H} = \chi - \chi_H$.

| Term index $i$ | Coefficient $c_i$ | Z-scored coefficient $z(c_i)$ | Term $t_i$ |
|---|---|---|---|
| 1 | $8.65 \times 10^{-1}$ | 0.498 | $\langle \Omega \rangle R_{Tr}$ |
| 2 | $-1.17$ | $-0.409$ | $\cos\left(\dfrac{\pi}{200} \dfrac{\langle V_p \rangle}{R_{Tr+RE}}\right)$ |
| 3 | $-1.92 \times 10^{-1}$ | $-0.225$ | $10^{\frac{\langle B \rangle}{\langle \theta_D \rangle}}$ |
| 4 | $-8.93$ | $-0.221$ | $\sin\left(\dfrac{\pi}{10} \langle \chi_{\Delta H} \rangle \langle r_M \rangle\right) \left(\mathrm{erf}\left(\langle v \rangle R_{Tr(IV)} - \dfrac{1}{20}\right)\right)^{-\frac{1}{2}}$ |
| 5 | $1.40$ | $0.198$ | $\left(\mathrm{erf}\left(100 \dfrac{\langle \eta_{fM} \rangle}{\langle V_p \rangle}\right)\right)^{-\frac{1}{2}}$ |
| 6 | $9.71 \times 10^{-1}$ | $0.170$ | $\cos\left(\dfrac{\pi}{200} \langle \kappa \rangle \langle \Omega_\sigma \rangle\right)$ |
| 7 | $3.84 \times 10^{-16}$ | $0.161$ | $\langle \Omega \rangle \langle V_p \rangle \langle V_p \rangle \dfrac{\langle \rho_{mol} \rangle \langle \kappa \rangle}{\left(10^{\sin\left(\frac{\pi}{1000}\frac{\langle r_M \rangle}{R_{Tr}}\right)} \mathrm{erf}(\langle G \rangle \langle \alpha \rangle - \frac{1}{20})\right)^{-\frac{1}{2}}} \times \dfrac{\langle \theta_D \rangle R_{Tr}}{\frac{\langle \alpha \rangle}{\langle \Omega \rangle} \langle \rho_{mol} \rangle \langle \kappa \rangle} \dfrac{\langle G \rangle \langle \kappa \rangle}{\exp(\sigma(\chi_{\Delta H}))} \times \mathrm{erf}(100 r(\alpha)) \left(\dfrac{\langle \rho_{mol} \rangle}{\langle G \rangle \langle \kappa \rangle \frac{\langle \rho \rangle}{\langle \rho_{mol} \rangle} \left(\log_{10}\left(\frac{\langle n_{ve} \rangle}{\langle G \rangle}\right)\right)^{-\frac{1}{2}}}\right)^{-\frac{1}{2}} (\exp(r(\chi_m)))^{-\frac{1}{2}}$ |
| 8 | $4.29 \times 10^{-6}$ | $0.121$ | $\langle B \rangle \langle \theta_D \rangle$ |
| 9 | $2.08 \times 10^{-1}$ | $0.117$ | $\sin\left(\dfrac{\pi}{10} \langle \chi_{\Delta H} \rangle \langle r_M \rangle\right) \sigma(\rho)$ |
| 10 | $-3.27 \times 10^{-1}$ | $-0.111$ | $\left(10^{\left(\mathrm{erf}(\langle v \rangle R_{Tr(IV)} - \frac{1}{20})\right)^{-\frac{1}{2}} \left(\mathrm{erf}\left(100(\log_{10}(\langle \rho \rangle \langle v \rangle))^{-\frac{1}{2}} \sin\left(\frac{\pi}{1000}\sigma(\theta_D)\right)\right)\right)^{-\frac{1}{2}}}\right)^{-\frac{1}{2}}$ |

| 11 | -2.43×10⁻⁶ | -0.0978 | $\left(\log_{10}\left(\frac{\langle M \rangle}{R_{\text{Tr}}}\right)\right)^{-\frac{1}{2}} \langle n_{ve} \rangle \langle \theta_D \rangle \cos\left(\frac{1}{1000} \langle M \rangle \langle \theta_D \rangle\right)$ |
|---|---|---|---|
| 12 | 4.29×10⁻² | 0.0793 | $\cos\left(\pi \langle \chi_{\Delta H} \rangle \langle \rho \rangle\right)$ |
| Intercept | -3.34 | | |

**Supplementary Table 1** provides a representative symbolic regression model for $\log_{10}(w/\langle M \rangle)$. The terms are listed in descending order of the magnitude of their $z$-scored coefficients, so that the relative importance of each contribution to the model can be directly assessed. This ordering helps visualize which descriptor combinations most strongly govern the predicted hydrogen capacity.

## b. Symbolic Model for $\log_{10} P_{eq,RT}$

**Supplementary Table 2** An example of symbolic model for $\log_{10} P_{eq,RT}$. Here, $\chi_{\Delta H} = \chi - \chi_H$.

| Term index $i$ | Coefficient $c_i$ | Z-scored coefficient $z(c_i)$ | Term $t_i$ |
|---|---|---|---|
| 1 | $5.13 \times 10^{-1}$ | 3.841 | $\langle G \rangle R_{\text{Tr+RE}}$ |
| 2 | $1.19 \times 10^{9}$ | 1.929 | $\sin\left(\frac{\pi}{20} \frac{\langle Z \rangle}{\langle r_M \rangle} \frac{\langle \alpha \rangle}{\langle \theta_D \rangle}\right)$ |
| 3 | $-8.68 \times 10^{-5}$ | -1.778 | $\dfrac{\langle G \rangle \langle \theta_D \rangle}{\frac{\langle v \rangle}{R_{\text{Tr+RE}}}}$ |
| 4 | $1.76 \times 10^{2}$ | 1.490 | $\dfrac{\langle \rho_{\text{mol}} \rangle}{R_{\text{Tr+RE}}}$ |
| 5 | $7.22 \times 10^{10}$ | 1.328 | $\left(\exp\left(\dfrac{\frac{\langle \alpha \rangle}{\langle \theta_D \rangle}}{\frac{\langle M \rangle}{\langle \rho_{\text{mol}} \rangle}}\right)\right)^{-\frac{1}{2}}$ |
| 6 | $-6.18 \times 10^{1}$ | -1.001 | $\sin\left(\frac{\pi}{2000} \langle B \rangle\right)$ |
| 7 | $-9.58 \times 10^{3}$ | -0.898 | $\langle G \rangle \langle \alpha \rangle$ |
| 8 | $-2.39$ | -0.779 | $(\log_{10}(\langle M \rangle \langle Z \rangle))^{-\frac{1}{2}}$ |
| 9 | $-3.85 \times 10^{-2}$ | -0.750 | $\sigma(\theta_D)$ |
| 10 | $1.33 \times 10^{-6}$ | 0.705 | $\langle n_{ve} \rangle \langle B \rangle \dfrac{\langle \kappa \rangle}{\langle \Omega_\sigma \rangle}$ |
| 11 | $9.78 \times 10^{-4}$ | 0.468 | $\langle \chi_{\Delta H} \rangle \langle M \rangle \dfrac{\langle r_M \rangle}{\langle \Omega_\sigma \rangle}$ |
| 12 | $1.57$ | 0.456 | $\sin\left(\frac{\pi}{20} \text{erf}\left(\frac{1}{10} r(B)\right) \langle \chi_{\Delta H} \rangle \langle M \rangle\right)$ |
| 13 | $-3.92 \times 10^{-2}$ | -0.379 | $\langle G \rangle R_{\text{Tr(IV)}}$ |
| 14 | $-1.17$ | -0.288 | $\sigma(Z)$ |
| 15 | $9.06 \times 10^{-2}$ | 0.213 | $\dfrac{\langle Z \rangle}{\langle r_M \rangle} r(\chi_{\Delta H})$ |
| 16 | $-1.09 \times 10^{-2}$ | -0.204 | $\langle \chi_{\Delta H} \rangle \langle M \rangle \sigma(\rho)$ |
| 17 | $-4.75 \times 10^{-3}$ | -0.204 | $\langle \kappa \rangle r(n_{ve})$ |
| 18 | $6.72 \times 10^{-2}$ | 0.194 | $(\sigma(n_c))^{-\frac{1}{2}}$ |
| 19 | $1.18 \times 10^{-1}$ | 0.0956 | $r(\kappa)$ |

| Intercept | -7.22×10^10 | | |

**Supplementary Table 2** provides a representative symbolic regression model for $\log_{10} P_{\text{eq,RT}}$. The terms are listed in descending order of the magnitude of their *z*-scored coefficients, so that the relative importance of each contribution to the model can be directly assessed. This ordering helps visualize which descriptor combinations most strongly govern the predicted hydrogen capacity.

## c. Error Distributions of Stacking-ensembled Models for $\log_{10}(w/\langle M \rangle)$ and $\log_{10} P_{eq,RT}$

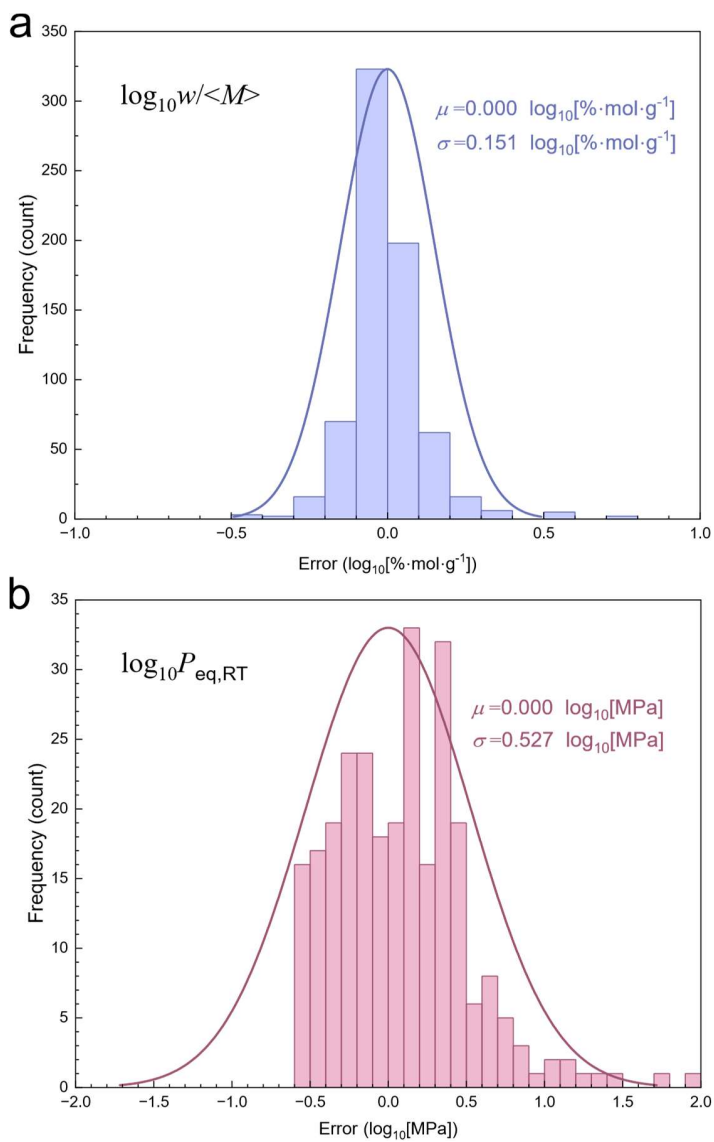

**Supplementary Figure 4** Error distributions of stacking-ensembled models for $\log_{10}(w/\langle M \rangle)$ and $\log_{10} P_{eq,RT}$. Error distributions for (a) $\log_{10}(w/\langle M \rangle)$ and (b) $\log_{10}(w/\langle M \rangle)$ obtained from the stacking-ensembled symbolic regression models. Solid curves represent fitted normal distributions, with the mean ($\mu$) and standard deviation ($\sigma$) indicated in each panel.

Both error distributions are centered around zero, indicating the absence of systematic bias in the models (see **Supplementary Figure 4**). These results confirm that the models provide reliable and unbiased predictions across the dataset.

**d. Stacking-ensembled Symbolic Models for $\log_{10} w$ and $\log_{10}(w\langle M \rangle)$**

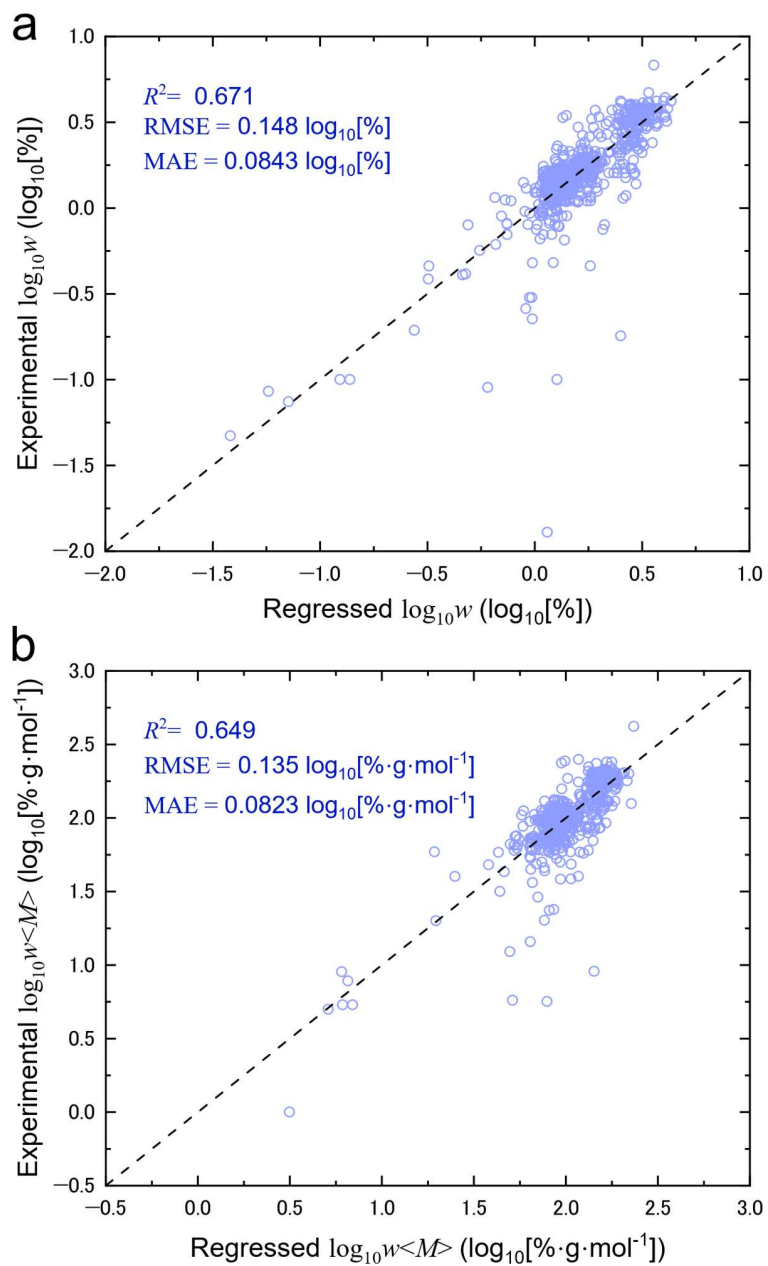

**Supplementary Figure 5** Stacking-ensembled symbolic models for $\log_{10} w$ and $\log_{10}(w\langle M \rangle)$. Parity plots for (a) $\log_{10} w$ and (b) $\log_{10}(w\langle M \rangle)$ obtained using stacking-ensembled symbolic regression models. Model performance metrics ($R^2$, RMSE, and MAE) are indicated in each panel.

Compared to the model using $\log_{10}(w/\langle M \rangle)$, both $\log_{10} w$ and $\log_{10}(w\langle M \rangle)$ exhibit lower predictive performance, as reflected by reduced $R^2$ values and broader scatter in the parity plots (see **Supplementary Figure 5**). This improvement reflects the differing roles of $\langle M \rangle$ across target definitions; whereas $w\langle M \rangle$ penalizes heavy-element systems and $w\langle M \rangle$ favors them, $w/\langle M \rangle$ suppresses mass dependence, thereby enabling clearer extraction of intrinsic structure-property relationships. This difference highlights the strong influence of $\langle M \rangle$ on these target definitions, which can obscure underlying structure-property relationships. As a result, normalization by $\langle M \rangle$ in the form of $w/\langle M \rangle$ provides a more effective representation for data-driven modeling.

**Supplementary References**

S1. Jang, S.-H. GoodRegressor: A hierarchical inductive bias for navigating high-dimensional compositional space. *arXiv*, February 20, 2026. DOI: 10.48550/arXiv.2510.18325

S2. Hilt, D. E.; Seegrist, D. W. *Ridge, a computer program for calculating ridge regression estimates*; Dept. of Agriculture, Forest Service, Northeastern Forest Experiment Station, 1977.

S3. Zou, H.; Hastie, T.; Tibshirani, R. Sparse principal component analysis. *J. Comput. Graph. Stat.* **2006,** *15* (2), 265-286. DOI: 10.1198/106186006X113430. 265

S4. Cybenko, G. Approximation by superpositions of a sigmoidal function. *Math. Control Signal Systems* **1989,** *2*, 303-314. DOI: 10.1007/BF02551274

S5. Liaw, A.; Wiener, M. Classification and regression by randomForest. *R News* **2002,** *2*, 18-22.

S6. Chen, T.; Guestrin, C. XGBoost: A scalable tree boosting system. *KDD '16: Proceedings of the 22nd ACM SIGKDD International Conference on Knowledge Discovery and Data Mining*, San Francisco, California, USA, August 13-17, 2016; Krishnapuram, B.; Shah, M.; Smola, A.; Aggarwal, C.; Shen, D.; Rastogi, R., Eds.; Association for Computing Machinery: New York, NY, United States, 2017; pp 785-794. DOI: 10.1145/2939672.2939785

S7. Izu, T. *et al*. LightGBM: A highly efficient gradient boosting decision tree. *NIPS'17: Proceedings of the 31st International Conference on Neural Information Processing Systems*, Long Beach, California, USA, December 4-9, 2017; von Luxburg, U.; Guyon, I.; Bengio, S.; Wallach, H.; Fergus, R. Eds.; Curran Associates Inc., Red Hook, New York, United States, 2017; pp 3149-3157. DOI: 10.5555/3294996.3295074

S8. Kim, S.; Lu, P, Y.; Mukherjee, S.; Gilbert, M.; Jing, L.; Čeperić, V. Integration of neural network-based symbolic regression in deep learning for scientific discovery. *IEEE. Trans. Neural. Netw. Learn. Syst.* **2021,** *32* (9), 4166-4177. DOI: 10.1109/TNNLS.2020.3017010

S9. Tenachi, W.; Ibata, R.; Diakogiannis, F. I. Deep symbolic regression for physics guided by units constraints: Toward the automated discovery of physical laws. *ApJ.* **2023,** *959* (2), 99. DOI: 10.3847/1538-4357/ad014c